\documentclass[%
 reprint,
 superscriptaddress,
 amsmath,amssymb,
 aps,
 prl,
floatfix,
]{revtex4-2}

\usepackage{graphicx}
\usepackage{booktabs}
\usepackage{nicefrac}
\usepackage{dcolumn}
\usepackage{bm}
\usepackage{braket}
\usepackage{chemformula}%

\usepackage{hyperref}
\usepackage{textcomp}

\newcommand{\HeThreeStar}{\ch{^3 He}*} 
\newcommand{\HeFourStar}{\ch{^4 He}*}

\newcommand{\avg}[1]{\left\langle #1 \right\rangle}

\begin{document}

\preprint{APS/123-QED}

\title{$n$-body anti-bunching in a degenerate Fermi gas of \HeThreeStar{} atoms}

\author{Kieran F. Thomas}
\altaffiliation{%
Present Address: Q-CTRL, Sydney, NSW, Australia
}%
\affiliation{%
 Department of Quantum Science and Technology, Research School of Physics, The Australian National University, Canberra, ACT 2601, Australia
}%

\author{Shijie Li}%
\affiliation{%
 Department of Quantum Science and Technology, Research School of Physics, The Australian National University, Canberra, ACT 2601, Australia
}%

\author{A. H. Abbas}
\altaffiliation{%
Present Address: Artificial Intelligence and Cyber Futures Institute, Charles Sturt University, Bathurst, NSW 2795, Australia
}%
\affiliation{%
 Department of Quantum Science and Technology, Research School of Physics, The Australian National University, Canberra, ACT 2601, Australia
}%
\affiliation{%
Department of Physics, Faculty of Science, Cairo University, Giza 12613, Egypt
}%

\author{Andrew G. Truscott}
\affiliation{%
 Department of Quantum Science and Technology, Research School of Physics, The Australian National University, Canberra, ACT 2601, Australia
}%

\author{Sean. S. Hodgman}
\email{sean.hodgman@anu.edu.au}
\affiliation{%
 Department of Quantum Science and Technology, Research School of Physics, The Australian National University, Canberra, ACT 2601, Australia
}%

\date{\today}

\begin{abstract}

A key observable in investigations into quantum systems are the $n$-body correlation functions, which provide a powerful tool for experimentally determining coherence and directly probing the many-body wavefunction. While the (bosonic) correlations of photonic systems are well explored, the correlations present in matter-wave systems, particularly for fermionic atoms, are still an emerging field. In this work, we use the unique single-atom detection properties of \HeThreeStar{} atoms to perform simultaneous measurements of the $n$-body quantum correlations, up to the fifth-order, of a degenerate Fermi gas. In a direct demonstration of the Pauli exclusion principle, we observe clear anti-bunching at all orders and find good agreement with predicted correlation volumes. Our results pave the way for using correlation functions to probe some of the rich physics associated with fermionic systems, such as $d$-wave pairing in superconductors.

\end{abstract}

\maketitle

Pair correlations for photons were first considered in attempts to explain the Hanbury-Brown and Twiss (HBT) effect,  where correlations between intensity fluctuations were observed for a thermal light source. 
\cite{hbt}.  The  increased probability of two photons arriving at a detector simultaneously compared to random chance (termed bunching) is caused by the constructive interference between individual photons.
Glauber famously reconciled this effect with a full quantum description of coherence based on $n$-body correlation functions, which started the field of quantum optics \cite{Glauber1963}. The experimental realisation of trapped neutral atoms at ultracold temperatures with large de Broglie wavelengths opened the possibility of conducting equivalent optics experiments with atoms rather than photons, termed quantum atom optics.  There have since been a number of studies on the measurement of $n$th-order bosonic correlation functions in a variety of ultracold systems \cite{hbt1, Schellekens2005, ttl2005,hbt2,Perrin2007,hbt3,s1,bs2,Hodgman2017,Carcy2019,cayla2020,A}. These explorations are significant for investigating quantum statistics, quantifying information on the coherence and size of a quantum source as well as providing deeper insight into many-body quantum behaviour.

\begin{figure}[b]
\includegraphics{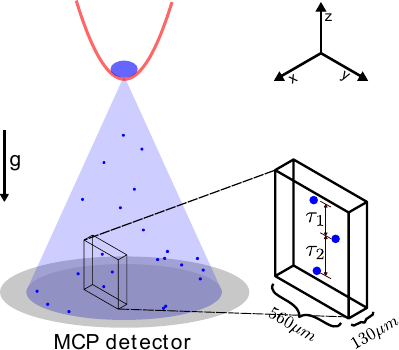}
\caption{Schematic showing how correlation functions are measured experimentally. An initially trapped cloud (top, blue) is released and expands as it falls onto an MCP detector (grey), which measures the arrival times of each atom as well as their x-y spatial locations. To construct the unnormalised third-order correlation function  ($G^{(3)}(\tau_1,\tau_2)$), all subsequent atoms arriving within a spatial volume $\Delta x$ and $\Delta y$ of each atom have their arrival time differences $\tau_1$ and $\tau_2$ (bottom right) recorded and histogrammed.}
\label{sc}
\end{figure}

Ultracold atoms also open up the fascinating possibility of measuring fermionic correlation functions, which have no equivalent in classical optics. In the HBT effect for fermionic fields, the anti-symmetry of the wavefunction leads to destructive interference between possible propagation paths and thus a decreased probability of simultaneous detections \cite{Jeltes2007}. This direct demonstration of the Pauli exclusion principle is referred to as anti-bunching and, unlike bosonic bunching, has no classical counterpart \cite{Rom2006}. However, 
partly due to there being less ultarcold fermionic experiments compared to bosons \cite{dfg}, there have been only a handful of studies on the second-order correlation function of neutral fermions \cite{Rom2006,Jeltes2007,Bergschneider2019} and a single measurement of 3 atom correlations \cite{PhysRevLett.122.143602}, but nothing on higher-order fermionic correlation functions. 

\begin{figure}[t]
\includegraphics{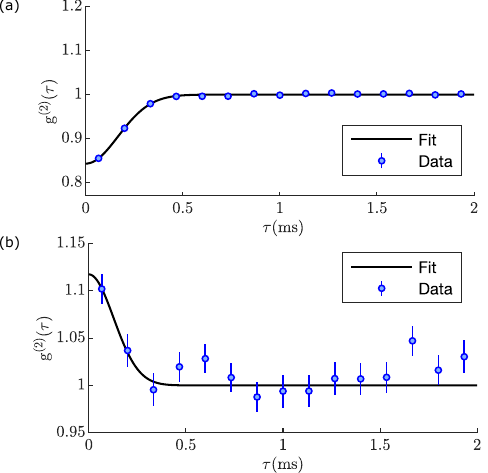}
\caption{The second-order normalised correlation function $g^{(2)}(\tau)$ for (a) fermionic \HeThreeStar{} atoms and 
 (b) thermal bosonic \HeFourStar{} atoms. The measured correlation amplitudes for (a) and (b) are $g^{(2)}(0)=0.84(1)$ and $g^{(2)}(0)=1.11(6)$, while the correlation lengths are $l_t = 240(10)$~$\mu$s and $l_t = 180(40)$~$\mu$s, respectively. The data in (a) has $\Delta t = 133$~$\mu$s, $\Delta x=130 \, \mu$m and $\Delta y = 560 \, \mu$m, and is averaged over 2,000 experimental runs. The data in (b) uses bin widths of $\Delta t = 100$~$\mu$s, $\Delta x=130 \, \mu$m and $\Delta y = 420 \, \mu$m and is averaged over 300 experimental runs, with each run containing 116 separately out-coupled clouds of atoms, similar to \cite{Manning2010}. The errors are estimated from the standard deviation of the counts of correlated tuples across all experimental runs.}
\label{f1}
\end{figure}

Here we present the first measurement of the normalised fermionic correlation functions simultaneously from second to fifth order for a ballistically expanding degenerate Fermi gas (DFG) of \HeThreeStar{} atoms. For comparison, we also present the second-order correlation function of a thermal cloud of \HeFourStar{} atoms, which is used to sympathetically cool the \HeThreeStar{} atoms, as they each display distinct behaviour \cite{Naraschewski1999,fermitheory,Gomes2006}. We utilise a Multi-Channel Plate (MCP) with a delay-line detector (DLD) to reconstruct the distribution of an ultracold \HeThreeStar{} and \HeFourStar{} mixture \cite{Thomas2023} with single atom resolution in the far field after ballistic expansion \cite{Manning2010}. This allows us to directly calculate correlation functions for both bosonic (\HeFourStar{}) and fermionic (\HeThreeStar{}) atoms, with our only limiting factor on maximum correlation function order achievable being data rates and detector resolution.
We are able to observe fermionic antibunching for every order up to $n=5$. We also investigate how the correlation width of the fermionic cloud varies with cloud size. Our results agree with theory \cite{Naraschewski1999,SOMs} when considering the effects of finite detector resolution and binning size.

Correlation functions were introduced by Glauber \cite{Glauber1963} to characterise the coherence between an $n$-tuple of particles in space and time. We consider an $n$th order correlation function that considers $n$-fold coincidence count rates
\begin{equation}
\begin{aligned}    &G^{(n)}\left(\textbf{r}_1,t_1;\ldots;\textbf{r}_n,t_n\right) 
    \\&\hspace{0.5cm}= \avg{\hat{\Psi}^\dagger(\textbf{r}_1,t_1) \hat{\Psi}(\textbf{r}_1,t_1) \ldots \hat{\Psi}^\dagger(\textbf{r}_n,t_n) \hat{\Psi}(\textbf{r}_n,t_n)},
\end{aligned}
\end{equation}
where $\hat{\Psi}^\dagger(\textbf{r}_i,t_i)$ is the field operator for a particle at position $\textbf{r}_i$ at time $t_i$ and the angle brackets denote averaging. Since here we will only be studying equilibrium distributions, we can ignore the $t_i$ variable and only consider $G^{(n)}\left(\textbf{r}_1;\ldots;\textbf{r}_n\right)$. To physically interpret the $n$th order correlation function, we take the normalised version
\begin{align}
    g^{(n)}(\textbf{r}_1;\ldots;\textbf{r}_n) &= \frac{G^{(n)}(\textbf{r}_1;\ldots;\textbf{r}_n)}{\rho(\textbf{r}_1) \rho(\textbf{r}_2) \ldots \rho(\textbf{r}_n)},
\end{align}
where $\rho(\textbf{r}_i)=G^{(1)}(\textbf{r}_i;\textbf{r}_i)=\avg{\hat{\Psi}^\dagger(\textbf{r}_i) \hat{\Psi}(\textbf{r}_i)}$, which gives the probability of detecting $n$ particles at points $(\textbf{r}_1)$ to $(\textbf{r}_n)$ with respect to random chance. For instance, a measurement of $g^{(n)}(0,\ldots,0)=1$ implies that the detection between the various points follows an uncorrelated Poissonian distribution. In contrast, $g^{(n)}(0,\ldots,0)>1$ implies there is bunching present in the system and $g^{(n)}(0,\ldots,0)<1$ implies anti-bunching. 

Our experiment starts with a combined degenerate Fermi gas of \HeThreeStar{} atoms and a Bose-Einstein condensate (BEC) of \HeFourStar{} atoms, both trapped in a magnetic trap with trapping frequencies for the \HeThreeStar{} atoms of $\omega_{x,y,z}=2\pi\times \left(58(3),694(1),701(2)\right)$Hz \cite{Thomas2023, Henson2022}.  After the trap switches off, the clouds fall $\sim$850mm (fall time $t_{TOF}$ = 416ms) onto a MCP and DLD, which can detect the 3D location of individual atoms with $\sim$130$\mu$m $x,y$ and $\sim$3$\mu$s $z$ resolution \cite{Manning2010,Henson2018}.  Due to the different masses of the two helium species, applying a small magnetic field gradient during time-of-flight (TOF) separates the arrival times of the clouds, allowing the distribution of each species to be measured in a single shot \cite{Thomas2023}.  The TOF is sufficiently long that the detected position distribution at the detector is close to the cloud's in-trap momentum distribution and hence the correlation functions correspond approximately to the in-trap momentum.

\begin{figure}[t]
    \centering
    \includegraphics{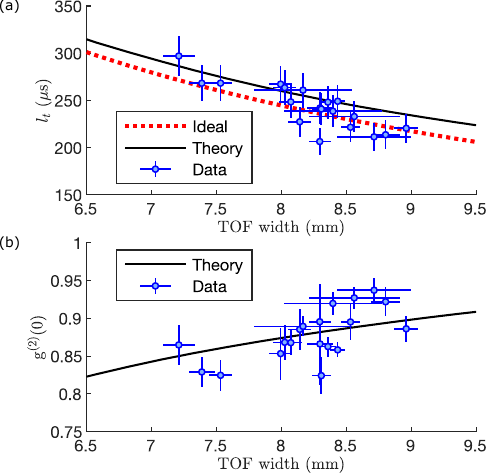}
    \caption{The experimentally measured temporal correlation length $l_t$ (a) and correlation amplitude $g^{(2)}(0)$ (b) for various time-of-flight widths of the \HeThreeStar{} cloud.  These are compared to the theoretically ideal value (dashed line) and expected value taking into account experimental factors (solid line) where realistic effects such as finite binning, resolution and dark counts are included via a Monte Carlo simulation \cite{SOMs}. For (b), the theoretically ideal value is $g^{(2)}(0)=0$ for all cloud widths, and hence it is not included in the plot. We find excellent agreement for both correlation length and amplitude between the experimentally measured values and the expected non-ideal values.}
    \label{fig:cloud_size}
\end{figure}

\begin{figure}[t]
\includegraphics{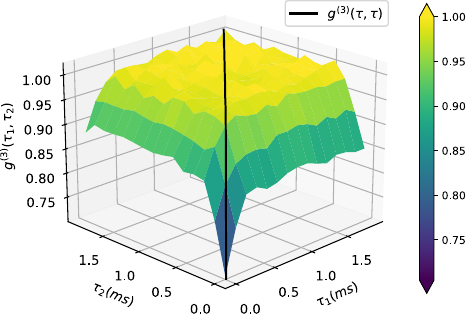}
\caption{Surface plot of the third-order normalised correlation function of a DFG of \HeThreeStar{} atoms $g^{(3)}(\tau_1,\tau_2)$. The data has $\Delta t = 133$~$\mu$s, $\Delta x=130 \, \mu$m and $\Delta y = 560 \, \mu$m, and is averaged over 2,000 experimental runs, each run corresponding to a single shot.}
\label{f1_2}
\end{figure}

The single atom resolution in 3D allows us to reconstruct correlation functions directly. For simplicity, we only consider correlations along the single axis with the highest detector resolution (the $z$ axis of the ballistic expansion). Since this corresponds to the arrival time of a falling cloud at the detector, we refer to this as correlations in arrival time $t$. We can therefore consider the simplified volume integrated correlation function, which for the second order is
\begin{align}
    g^{(2)}(\tau) &\equiv  \frac{\iint d\textbf{r} \, dt \, G^{(2)}(\textbf{r},t;\textbf{r},t+\tau)}{\iint d\textbf{r} \, dt \rho(\textbf{r},t) \rho(\textbf{r},t+\tau) }. 
\end{align}
Here, we have averaged over all arrival times and are hence now finding the correlation between detected events with a $\tau$ delay between their arrival times. The area of the spatial integral over the $x$ and $y$ directions is chosen to be comparable to the correlation lengths in these directions so that all events within the integration volume are correlated. We will also extend this logic to the $n$th order correlation to obtain the simplified volume integrated correlation function $g^{(n)}(\tau_1,\ldots,\tau_{n-1})$, as shown in Fig. \ref{sc} for $g^{(3)}(\tau_1,\tau_{2})$ (see \cite{SOMs} for further details). 
  
\begin{figure*}[t]
\center
\includegraphics{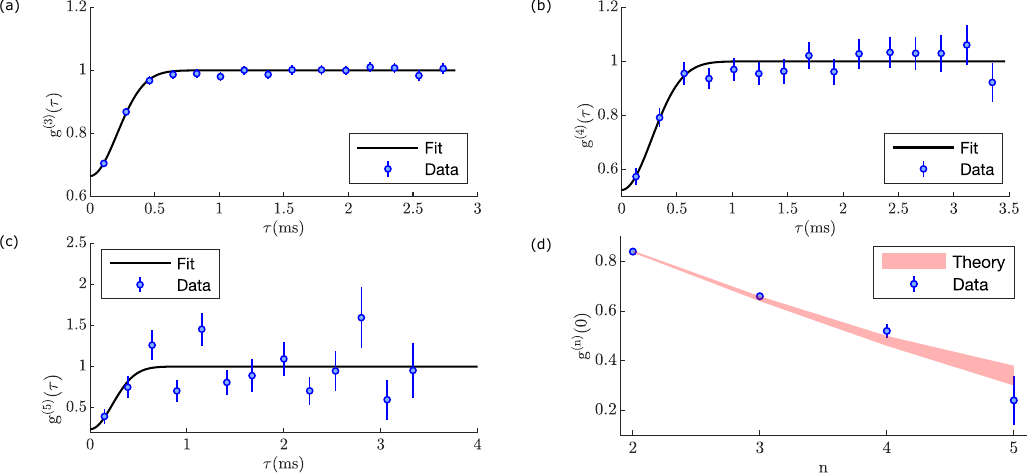}
\caption{The normalised correlation functions of a DFG of \HeThreeStar{} atoms at orders: (a) $g^{(3)}(\tau)$, (b) $g^{(4)}(\tau)$, and (c) $g^{(5)}(\tau)$. (d) The correlation amplitude, denoted as $g^{(n)}(0)$, for values of $n$ ranging from 2 to 5, is presented, with the red shaded area illustrating the theoretically predicted values of amplitude. The diagonal correlation lengths extracted from the fitted gaussians in (a), (b) and (c) are $300(10)$~$\mu s$, $390(50)$~$\mu s$ and $340(100)$~$\mu s$, respectively. All orders have bin sizes $\Delta t = 100$~$\mu$s, $\Delta x=130 \, \mu$m and $\Delta y = 420  \, \mu$m and are averaged over 2,000 experimental runs. The errors are estimated from the standard deviation of the counts of correlated tuples across all experimental runs.}
\label{f2}
\end{figure*}

 In Fig.~\ref{f1} (a) and (b), we show the experimentally measured $g^{(2)}(\tau)$ for a cloud of ultracold bosons (a) and fermions (b). Both distributions have the expected Gaussian form, with a bunching effect ($g^{(2)}(0)>1$) visible in the thermal \HeFourStar{} atoms and anti-bunching ($g^{(2)}(0)<1$) evident for the DFG of \HeThreeStar{} atoms. The temperature of the \HeFourStar{} atoms is $T=200(30)$~nK, determined by fitting the thermal wings of the \HeFourStar{} distribution, as in \cite{Thomas2023}. 
 
 By fitting Gaussians to the data in Fig.~\ref{f1} (a) and (b) \cite{SOMs}, we find the correlation length at the detector for the DFG is $l_t=240(10) \,\mu$s,  while for the thermal \HeFourStar{} cloud it is $l_t=180(40) \,\mu$s.
Incorporating experimental factors such as finite binning and detector resolution (as in previous investigations \cite{hbt2}), we perform a Monte Carlo simulation (see ~\cite{SOMs} for details) and find an expected correlation amplitude of $g^{(2)}(0)=0.83$ for the DFG, which agrees well with the experimentally measured value of $0.84(1)$. We similarly find the expected correlation length for the \HeThreeStar{} atoms $l_t=269$~$\mu$s, which again agrees well with the measured value of $270(10)$~$\mu$s. 

To further investigate the dependence of the anti-bunching on temperature, we vary the final temperature of the \HeThreeStar{} atoms by evaporating more of the coolant gas (bosonic \HeFourStar{}). This reduces the temperature while keeping the total number of fermions approximately the same. 
As there are few bosons left in the mixture and most of those remaining are in the condensate, it becomes difficult to ascertain an accurate measure of temperature from the \HeThreeStar{} cloud, 
as discussed in Ref.~\cite{Thomas2023}. Hence, we use the time-of-flight width of the Fermi cloud, given by $\sigma_{TOF}=\frac{t_{TOF}}{\gamma(\xi)} \sqrt{\frac{k_B T}{m}}$ (see \cite{SOMs} for details), as a proxy for temperature \footnote{The reason we do not use this expression to estimate temperature directly is that $\gamma(\xi)$ diverges at low temperatures, making a straightforward conversion very difficult. To make a proper estimate of temperature, the temperature $T$ and the degeneracy parameter $\gamma$ (which is itself temperature dependent) must be calculated together using a full fit of the TOF distribution and an atom number estimate. See \cite{Thomas2023} for further details.}. Note that the time-of-flight width is isotropic even for a degenerate Fermi cloud. Fig.~\ref{fig:cloud_size} shows the measured correlation length $l_t$ (a) and maximum anti-bunching amplitude $g^{(2)}(0)$ (b) as a function of $\sigma_{TOF}$, along with theoretical expected values \cite{SOMs}. 

Higher order correlation functions can also be calculated from the same dataset. The full representation of a higher-order correlation function requires a $n+1$ dimensional graph - for example $g^{(3)}(\tau_1,\tau_2)$ for our data is shown in the 3D surface plot in Fig.~\ref{f1_2}, with $(\tau_1,\tau_2)$ as defined in Fig.~\ref{sc}. The anti-bunching effect is visible in the reduced value of $g^{(3)}(\tau_1,\tau_2)$ as $\tau_1,\tau_2\rightarrow0$. As representations of higher-order correlation functions are difficult to visualise, we instead plot the diagonal correlation $g^{(n)}(\tau) \equiv g^{(n)}(\tau,...,\tau)$, where all differences are equal (i.e. $\tau_1=\tau_2...\tau_{n-1}=\tau$). The diagonal correlation function for $n=3$ is indicated by the black line in Fig.~\ref{f1_2}. In Fig.~\ref{f2}, we plot $g^{(n)}(\tau)$ for $n=3$ (a), $n=4$ (b) and $n=5$ (c). Each plot shows anti-bunching for $g^{(n)}(0)$, and we observe that the minimum anti-bunching amplitude decreases with increasing order, as shown in Fig.~\ref{f2} (d).  

To characterise this change in $g^{(n)}(0)$ with $n$, we assume that the atoms are non-interacting particles, meaning that we can apply Wick's theorem to decompose all higher-order correlation functions into a function of first-order correlation functions $g^{(1)}(\tau)$ \cite{s1,Gomes2006,g1_note}. The exact form of this decomposition differs for bosons and fermions. For example, the third-order correlation function can be expressed as \cite{SOMs}
\begin{align}
    g^{(3)}(\tau_1,&\tau_2) = 1+ 2 \mathcal{R} \left( g^{(1)}(\tau_1) g^{(1)}(\tau_2) g^{(1)}(\tau_2 + \tau_1)\right)\label{eqn:g3_g1}\\ & +\eta \left(|g^{(1)}(\tau_1)|^2 +|g^{(1)}(\tau_2)|^2 + |g^{(1)}(\tau_2+\tau_1)|^2\right) ,\nonumber 
\end{align}
where $\eta = -1$ for fermions and $+1$ for bosons. 
The general form of the decomposition of $g^{(n)}(0)$ in terms of $g^{(1)}(0)$ for fermions ($\eta=-1$) given by Wick's theorem \cite{SOMs} is
\begin{align}
    g^{(n)}(0) = 1-\sum_{k=2}^n \binom{n}{k} (k-1) (-1)^{k}  g^{(1)}(0)^k, \label{eqn:decomp_full2}
\end{align}
where $\binom{n}{k}$ is the binomial coefficient. 
Generally, $g^{(1)}(0)=1$, and hence all powers of $g^{(1)}(0)$ also equal 1, leading to $g^{(n)}(0)=0$ for all $n$ \footnote{To clarify this point further it is easier to consider the form $g^{(n)}(0) = \sum_{\nu\in P(n)} \eta^{i(\nu)}  g^{(1)}(0)^{|supp(\nu)|}$ (see \cite{SOMs} for derivation) where $P(n)$ is the set of permutations of the numbers 1 to $n$, $supp(\nu)$ is the support of the permutation $\nu$, and $i(\nu)$ is the number of inversions contained in the permutation. Given that there is an equal number of permutations with odd and even numbers of inversions, we will have an even number of positive negative, which will exactly cancel to give zero.}. 

However, for the finite experimental resolution and bin size used here, the amplitude of the anti-bunching will be significantly reduced, similar to what has been previously observed for bosonic bunching \cite{Gomes2006,hbt2}. Since the resolution and bin size are the same for all dimensions ($\tau_1,\dots,\tau_{n-1}$), the first-order correlation function is the same for all $\tau_i$, i.e. ($g^{(1)}(\tau_1 \rightarrow 0)=\ldots=g^{(1)}(\tau_{n-1} \rightarrow 0)$). Thus, the relation in Eqn.~\ref{eqn:decomp_full2} can be used to predict the experimentally expected value of higher order $g^{(n)}(\tau\rightarrow 0)$ when the experimentally determined value of $g^{(1)}(0)$ does not equal $1$, by using the value of $g^{(1)}(0)$ extracted from lower order experimentally measured correlation functions. Given this, we can extrapolate the predicted amplitudes of higher-order correlation functions based on the measured amplitude of $g^{(1)}(0)$ implied from lower-order ones. Based on the measured value of $g^{(2)}(0)=0.84(1)$ (see Fig.~\ref{f1}), we find $g^{(1)}(0)=0.40(1)$.  The amplitudes from third to fifth order are hence predicted to be $g^{(3)}\left(0\right)=0.65(1)$, $g^{(4)}\left(0\right)=0.48(2)$ and $g^{(5)}\left(0\right)=0.34(4)$, respectively. The shaded region in Fig.~\ref{f2}(d) compares this to the value measured from the actual correlation functions. These values all compare well with the experimental values of $g^{(3)}(0)=0.66(1)$, $g^{(4)}(0)=0.52(3)$ and $g^{(5)}(0)=0.3(1)$s.

In conclusion, we have measured the fermionic normalised correlation functions from second to fifth order, with orders greater than three being presented for the first time. The observed anti-bunching effect in $n$-tuples of \HeThreeStar{} atoms agrees with the predictions of Wick's theorem combined with realistic experimental effects.  Our method could be extended to in principle measure arbitrarily high-order correlations, with the major limiting factor being data acquisition rates.
The ability to measure higher-order correlation functions of fermionic atoms opens up a range of fascinating possibilities with \HeThreeStar{} atoms. Such experiments include probing interesting many-body phenomena such as time crystals \cite{Alaeian2022} and $d$-wave superconductors \cite{Kitagawa2011} or investigating foundational quantum mechanical effects such as the weak equivalence principle \cite{Geiger2018}.

\section*{Acknowledgements}

This work was supported through Australian Research Council Discovery Project Grant No. DP190103021. S.S.H. was supported by Australian Research Council Future Fellowship Grant No. FT220100670. K.F.T. was supported by an Australian Government Research Training Program scholarship.

\bibliography{Refs}

\begin{thebibliography}{35}
\bibitem [{g1_()}]{g1_note}%
  \BibitemOpen
  \href@noop {} {}\bibinfo {note} {Correlation function are normalised in such
  a way as to attain unit value in the case of perfect coherence. For the first
  order correlation function this means the normalisation differs slightly and
  is given by
  $g^{(1)}(r_1,t_1;r_2,t_2)=\frac{G^{(1)}(r_1,t_1;r_2,t_2)}{\sqrt{\rho(r_1,t_1)}\sqrt{\rho(r_2,t_2)}}$,
  with the volume integrated version given by $g^{(1)}(\tau)=\frac{\iint dr \,
  dt \, G^{(1)}(r,t;r,t+\tau)}{\iint dr \, dt \,
  \sqrt{\rho(r,t)}\sqrt{\rho(r,t+\tau)}}$.}\BibitemShut {Stop}%
\end{thebibliography}

\clearpage
\onecolumngrid

\section{Supplementary Material}

\subsection{Decomposition of $g^{(n)}$ via Wick's theorem}
\label{sec:wick}
The general form of Wick's theorem can be simplified for a correlation function using its respective anti-commutation or commutation relations (see example 5 in~\cite{bardenet2022point})
\begin{align}
    \avg{\hat{N}_1\ldots\hat{N}_n} &= \sum_{\nu\in P(n)} \eta^{i(\nu)} \prod_{j=1}^n \avg{\hat{\Psi}^\dagger_j \hat{\Psi}_{\nu(j)}}. \label{eqn:wicks_coherence}
\end{align}
Here $P([1,n])$ is the set of permutations of the numbers 1 to $n$, $i(\nu)$ is the number of inversions contained in the permutation $\nu$, $\hat{N}_k = \hat{\Psi}^\dagger_k \hat{\Psi}_{k}$ is the number operator, $\nu(j)$ represents the $j$th term in the permutation $\nu$ and $\eta$ is $+1$ for bosons and $-1$ for fermions. We can readily identify the LHS of Eqn.~\ref{eqn:wicks_coherence} as $G^{(n)}(\textbf{r}_1,t_1;\ldots;\textbf{r}_n,t_n)$ (for now we will consider $G^{(n)}(t_1;\ldots;t_n)$ for notational convenience and relevance, however, this derivation is also valid for an arbitrary position vector in some parameter space) and the RHS as a function of various $G^{(1)}$'s. Using the notation of time differences, we can then rewrite Eqn.~\ref{eqn:wicks_coherence} as
\begin{align}
    G^{(n)}(t;\ldots;t_{n}) &= \sum_{\nu\in P([1,n])} \eta^{i(\nu)} \prod_{j=1}^{n}  G^{(1)}(t_j;t_{\nu(j)}),
\end{align}
where $G^{(1)}(t_k;t_m)=\avg{\hat{\Psi}^\dagger_j \hat{\Psi}_{\nu(j)}}$ by definition. As we wish to consider the normalised correlation functions, we divide both sides by $\rho_1\ldots\rho_n$ to give
\begin{align}
    g^{(n)}(t_1,\ldots,t_{n}) &= \sum_{\nu\in P([1,n])} \eta^{i(\nu)} \prod_{j=1}^{n} \frac{G^{(1)}(t_j;t_{\nu(j)})}{\sqrt{\rho_j} \sqrt{\rho_{\nu(j)}}}.
\end{align}
Note that we have used the fact that each term in the sum has exactly one $\hat{\Psi}^\dagger_k$ and $\hat{\Psi}_{k}$ for each $k$ between 1 to $n$ and have `assigned' each of these a normalisation of $\sqrt{\rho_k}$ from $\rho_1\ldots\rho_n$. If $\nu(j)=j$ we have $G^{(1)}(t_j;t_j)=\rho_j$, hence by definition, these terms are cancelled by the denominator, otherwise we obtain $g^{(1)}(t_j;t_{\nu(j)})$. Thus, 
\begin{align}
    g^{(n)}(t_1,\ldots,t_{n}) &= \sum_{\nu\in P([1,n])} \eta^{i(\nu)} \prod_{j\in supp(\nu)} g^{(1)}(t_j,t_{\nu(j)}), \label{eqn:decomp_general}
\end{align}
where $supp(\nu)$ is the support of the permutation $\nu$ \footnote{The support of a permutation is the set of numbers (or more generally elements) that are moved from their original positions under that permutation. For example, the support of the permutation $1234\mapsto 2134$ is $\{1,2\}$.}. We can use Eqn.~\ref{eqn:decomp_general} to derive any order correlation function. Tab.~\ref{tab:gn} shows the correlation functions up to $n=5$. 

To match the measured correlation function, we simplify further by substituting $t_k=t+\tau_{k-1}$ and integrating over $t$ to obtain the averaged correlation function, 
\begin{align}
    g^{(n)}(\tau_1,\ldots,\tau_{n-1}) &= \sum_{\nu\in P([0,n-1])} \eta^{i(\nu)} \prod_{j\in supp(\nu)} \int dt \, g^{(1)}(t+\tau_j;t+\tau_{\nu(j)})
\end{align}
where we have shifted the indexation down by 1 to reflect the relabeling and $\tau_0=0$. Next notice $\int dt\, g^{(1)}(t+\tau_j;t+\tau_{\nu(j)})=\int dt'\, g^{(1)}(t';t'+\tau_{\nu(j)}-\tau_j)$, thus we can write these terms as $g^{(1)}(\tau_{\nu(j)}-\tau_j)$. From this, we see the exact form of the decomposition of the average $n$th-order correlation function, $g^{(n)}(\tau_1,\ldots,\tau_{n-1})$, in terms of $g^{(1)}(\tau)$'s, is given by Wick's theorem as
\begin{align}
    g^{(n)}(\tau_1,\ldots,\tau_{n-1}) = \sum_{\nu\in P([0,n-1])} \eta^{i(\nu)} \prod_{j\in supp(\nu)} g^{(1)}(\tau_{\nu(j)}-\tau_j).\label{eqn:decomp_full}
\end{align}

In the text we consider the diagonal correlation function $g^{(n)}(\tau) \equiv g^{(n)}(\tau,...,\tau)$, and specifically we want to understand the behaviour of $g^{(n)}(\tau \rightarrow 0)$. 
Note $\tau$ can be positive or negative, however, we assume the same sign for all particles (i.e. $\tau_1,\ldots,\tau_{n-1}$ all have the same sign). In the ideal case of $g^{(1)}(\tau \rightarrow 0)=1$, Eqn.~\ref{eqn:decomp_full} gives the interesting result that if $n\geq2$ and $\eta=-1$ then $g^{(n)}(\tau)=0$ for all $\tau$. This intuitively makes sense, as $g^{(n)}(\tau)>0$ for $n\geq2$ would require two fermions to be measured at the same location, which in the ideal case is impossible. We will see, however, there is deviation from this ideal case experimentally due to various effects. Hence, to predict the value of $g^{(n)}(0)$ under realistic measurement conditions we set $\tau=0$ (assuming nothing of the value of $g^{(n)}(0)$) and reach 
\begin{align}
    g^{(n)}(0) = \sum_{\nu\in P([0,n-1])} \eta^{i(\nu)}  g^{(1)}(0)^{|supp(\nu)|}. \label{eqn:decomp_comb}
\end{align}
This can be rewritten as
\begin{align}
    g^{(n)}(0) = 1+\sum_{k=2}^n T(k) \binom{n}{k} \eta^{k-1}  g^{(1)}(0)^k, \label{eqn:decomp_zero_simp}
\end{align}
where $T(k)= !k$ for bosons ($\eta=1$), with $!k$ denoting the sub-factorial of $k$, and $T(k)= (k-1)$ for fermions ($\eta=-1$). To understand this simplification consider that for $|supp(\nu)|=k$, we must displace $k$ elements of our original set of $n$, so we $\binom{n}{k}$ choices of the set of elements that we are displacing. For bosons we then simply consider the number of possible ways we can permute these $k$ elements such that none are left in their original positions, known as a derangement, of which there are $!k$ ways to do for such a number of elements. For fermions, we instead need to consider how many derangements have a different parity of inversions. This is given by $(-1)^{k-1} (k-1)$, noting this difference is independent of the original set of elements we chose.

For $g^{(1)}(0)=1$ we obtain the expected relations of $g^{(n)}(0)=n!$ for bosons and $g^{(n)}(0)=0$ for fermions. For bosons, this is can be understood by noting that each term in the sum from Eqn.~\ref{eqn:decomp_comb} contributes 1, and thus the total sum equals the number of terms, i.e. the number of permutations of the numbers from $0$ to $n-1$, which is $n!$. For fermions, we see that the sum in Eqn.~\ref{eqn:decomp_comb} is equal to the difference between the number of permutations with odd and even numbers of inversions, which is always equal, and hence we obtain $0$. As discussed in the main text, this implies that the measured value of $g^{(n)}(\tau\rightarrow 0)$ can be predicted solely by the experimentally measured value $g^{(1)}(0)$. It can be seen that the measured value rapidly deviates from zero for $g^{(1)}(0)\neq 0$, explaining the measured $g^{(n)}(\tau)$ distributions (see main text Fig.~4).

\begin{table}[]
\renewcommand{\arraystretch}{1.5}
    \centering
    \caption{Decomposition of the $n$th order fermionic ($\eta=-1$) correlation functions, for $n$ from 2 to 5, in terms of the first-order correlation function. The notation $S^k$ represents the set of $k$ element subsets of the set $S$. Alternatively, this can be thought of as the set of possible outcomes for the operation $S$ choose $k$. We have also used the notation $\mathbb{D}_R(S)=\{k\in M(D(S)) : |k|=R\}$, i.e. the set of elements of $M(D(S))$ with cardinality equal to $R$, where $D$ are the derangements of the set $S$ \footnote[1]{Permutations of $S$ such that no element remains at the same place. This can also be stated as the support of the permutation is equal to the original set $S$.} 
    and $M$ acts on a set of permutations $X$ as follows 
    $M(X)=\Big\{\big\{ \{i,\nu(i)\}: i \in [1,|\nu|]\big\} : \nu \in X \Big\}$, giving the mapping 
    set \footnote[2]{This can be thought of as the set of the sets of pairs of elements that map to each other for each permutation contained in set $X$.}. Generally, $t_i$ can be thought of as some arbitrary dimension position vector. For comparison, we show the simplified decomposition of $g^{(n)}(0)$, where all $t_i=0$.}
    \begin{tabular}{c|ll}
    \toprule
    \toprule
       $n$ & $g^{(n)}(t_1,\ldots,t_n)$ \footnote[4]{In this column of the table, we have used the indexation shorthand $(i,j) \in S$ for $\{\{i,j\}:\,\{i,j\}\in S$ and $i<j\}$. The purpose of the implicit condition $i<j$ is to disambiguate the indexation as if $\{i,j\}\in S$ then $\{j,i\}\in S$. Thus if a sum or product over $\{i,j\}\in S$ were followed strictly it could lead to double counting.} &  $g^{(n)}(0)$  \\
       \hline
        2 & $1-\left|g^{(1)}(t_1,t_2)\right|^2$ & $1-g^{(1)}(0)^2$\\
        3 & $\begin{aligned}1-\sum_{(i,j) \in [1,3]^2} \left|g^{(1)}(t_{i},t_{j})\right|^2+2R\Big(g^{(1)}(t_1,t_2) g^{(1)}(t_1,t_3) g^{(1)}(t_2,t_3)\Big)\end{aligned}$ & $1-3g^{(1)}(0)^2+2g^{(1)}(0)^3$\\
        4 & $\!\begin{aligned}[t]1-&\sum_{(i,j) \in [1,4]^2} \left|g^{(1)}(t_{i},t_{j})\right|^2+\sum_{k\in[1,4]^3} 2R\Big(\prod_{(i,j)\in k^2} g^{(1)}(t_{i},t_{j})\Big) + \\
        &\sum_{k\in \mathbb{D}_2([1,4]) } \prod_{(i,j) \in k} \left|g^{(1)}(t_{i},t_{j})\right|^2 - \sum_{k\in \mathbb{D}_4([1,4]) } 2 R\Big(\prod_{(i,j) \in k} g^{(1)}(t_{i},t_{j}) \Big) \end{aligned}$ & $1-6g^{(1)}(0)^2+8g^{(1)}(0)^3-3g^{(1)}(0)^4$ \\
        5 & $\!\begin{aligned}[t]1-&\sum_{(i,j) \in [1,5]^2} \left|g^{(1)}(t_{i},t_{j})\right|^2+\sum_{k\in[1,5]^3} 2R\Big(\prod_{(i,j)\in k^2} g^{(1)}(t_{i},t_{j})\Big) + \\
        &\sum_{l\in [1,5]^4} \left(\sum_{k\in \mathbb{D}_2(l) } \prod_{(i,j) \in k} \left|g^{(1)}(t_{i},t_{j})\right|^2 - \sum_{k\in \mathbb{D}_4(l) } 2 R\Big(\prod_{(i,j) \in k} g^{(1)}(t_{i},t_{j}) \Big) \right) \end{aligned}$ & $1-10g^{(1)}(0)^2+20g^{(1)}(0)^3-15g^{(1)}(0)^4+4g^{(1)}(0)^5$ \\ 
        \bottomrule
    \end{tabular}
    
    \label{tab:gn}
\end{table}

\subsection{Exact form of correlation length for non-interacting fermi gases in a harmonic trap}
\label{sec:exact}

In the text while we are measuring momentum correlations, we construct these correlations using a time-of-flight distribution and hence discuss correlators in terms of detected position and time, however for this derivation we will use the equilibrium position and momentum of the trapped atoms, i.e. the underlying distribution which produces the time-of-flight profile. We discuss how to convert these momentum correlation lengths to their time-of-flight equivalents in the text and Sec.~\ref{sec:predict}. 
\\
From Sec.~\ref{sec:wick}, we can see that if we find an analytical form for the correlation length of $G^{(1)}\left(\textbf{p}_1;\textbf{p}_2\right)$ we can use Wick's theorem to extend it to find an analytical form for any $G^{(n)}\left(\textbf{p}_1;\ldots;\textbf{p}_n\right)$. The definition of the first-order coherence function can be rewritten in terms of the Wigner function $W(\bf{p},\bf{q})$ \cite{Naraschewski1999},
\begin{align}
    G^{(1)}(\bf{r},\bf{r}') &= \int d \bf{p} e^{-i \bf{p} \cdot  (\bf{r}-\bf{r}')/\hbar } W(\bf{p},\frac{\bf{r}+\bf{r}'}{2})\\
    G^{(1)}(\bf{p},\bf{p}') &= \int d \bf{r} e^{-i \bf{r} \cdot  (\bf{p}-\bf{p}')/\hbar } W(\frac{\bf{p}+\bf{p}'}{2},\bf{r}), \label{eqn:g1_def}
\end{align}
with $\bf{r},\bf{r}'$ and $\bf{p},\bf{p}'$ referring to the in-trap position and momentum, respectively. Under the local density (or semi-classical) approximation, which assumes $W(\bf{p},\bf{q})$ is equivalent to a spatially homogeneous system with a varying chemical potential, the Wigner function of harmonically trapped atoms is given by
\begin{align}
    W(\bf{p},\bf{q}) &= \frac{1}{(2\pi \hbar)^3} \frac{1}{Exp\left[ \beta \bf{p}^2+\alpha \bf{q}^2\right]/\xi-\eta}, \label{eqn:wigner}
\end{align}
where $\alpha=\frac{m}{2k_BT}$, $\beta = \frac{1}{2m k_B T}$, $\bf{q}=(\omega_x x)^2+(\omega_y y)^2+(\omega_z z)^2$ is the normalised position vector, and $\xi=e^{\mu/k_BT}$ is the fugacity. Notice that exchanging the momentum $\bf{p}$ and position $\bf{q}$ in a harmonic trap is equivalent to exchanging the values of $\beta$ and $\alpha$. Thus, the momentum and position correlation functions are functionally equivalent for non-interacting gases in a harmonic trap.

The main quantity of interest measured in this work is the average normalised correlation function. We consider the full definition
\begin{align}
    g^{(1)}(\Delta p) &\equiv \frac{\int d\textbf{P}  \, G^{(1)}(\textbf{P}-\frac{\Delta p}{2};\textbf{P}+\frac{\Delta p}{2})}{\int d\textbf{P} \, \sqrt{\rho(\textbf{P}-\frac{\Delta p}{2})} \sqrt{\rho(\textbf{P}+\frac{\Delta p}{2})} },\label{eqn:g_1_def}
\end{align}
where $\rho(\textbf{p}_i)=G^{(1)}(\textbf{p}_i;\textbf{p}_i)$ is the momentum density function. Using Eqns.~\ref{eqn:g1_def} and \ref{eqn:wigner}, we can approximate the solution of Eqn.~\ref{eqn:g_1_def} (assuming $\Delta p \ll (2m k_B T)^{\frac{1}{2}}$) as

\begin{align}
    g^{(1)}(\Delta p) \sim \exp\left[-\frac{\Delta p^2}{2l_c^2} \right] \label{eqn:g1_gauss}
\end{align}
with correlation length $l_i = \frac{\hbar}{2s_i} \gamma(\xi)$, where $s_i=\sqrt{\frac{kT}{m\omega_i^2}}$ is the size of a thermal cloud in the $i$-axis of a harmonic trap and $\gamma(\xi)=\sqrt{\frac{\text{Li}_{3}(\eta \xi)}{\text{Li}_{4}(\eta \xi)}}$. This approximation accurately predicts the bulk behavior of the correlation functions, even in highly degenerate regimes. Accordingly, we can express $g^{(2)}(\Delta p)$ and $g^{(3)}(\Delta p_1,\Delta p_2)$ for fermions as Gaussian functions with corresponding correlation lengths over the whole temperature range of interest. Following from the relations in Tab.~\ref{tab:gn}, we find

\begin{align}
    g^{(2)}(\Delta p) &= 1- e^{-\frac{\Delta p^2}{l_c^2}}\label{eqn:g2_gauss}\\
    g^{(3)}(\Delta p_1,\Delta p_2) &= 1-e^{-\left(\frac{\Delta p_1}{l_c}\right)^2}- e^{-\left(\frac{\Delta p_2}{l_c}\right)^2}- e^{-\left(\frac{\Delta p_1+\Delta p_2}{l_c}\right)^2}+2e^{-\frac{\Delta p_1^2+\Delta p_2^2+(\Delta p_2+\Delta p_1)^2}{2l_c^2}}. 
\end{align}
Note that the sign of the $\Delta p_2+\Delta p_1$
terms is due to us considering an ordered set of events. If we were looking at an unordered set, we must consider an average of $\Delta p_2+\Delta p_1$ and $\pm(\Delta p_2-\Delta p_1)$.

\subsection{Predicted experimental correlation amplitudes and lengths}
\label{sec:predict}
\begin{figure}[t]
    \centering
    \includegraphics[width=\textwidth]{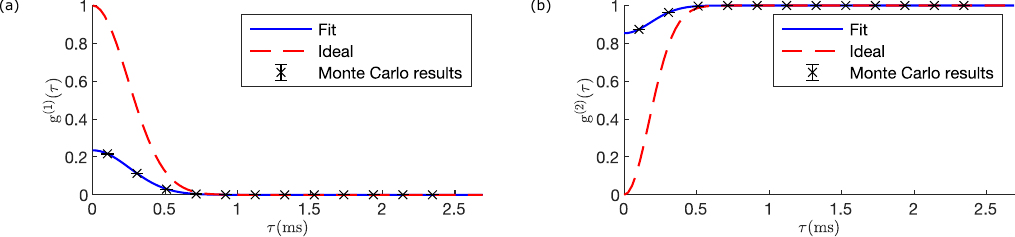}
    \caption{Results of the Monte Carlo simulation for (a) first and (b) second order Fermi correlation functions (black x's) for correlation lengths of $(l_t, \, l_x,\, l_y)$=$(244.6\,\mu\text{s},\, 83.0 \,\mu\text{m},\, 996.6 \,\mu\text{m})$ (corresponding to $\sigma_{TOF}=8$~mm). The predicted correlation length and amplitude for $g^{(2)}(\tau)$ are $l_t^{\text{SIM}}=250(1)$~$\mu$s and $g^{(2)}(0)_{\text{SIM}}=0.853(1)$ respectively.  These are extracted by fitting a gaussian to the simulation output (blue line). For comparison, we also show the ideal distributions, plotted using Eqns.~\ref{eqn:g1_gauss} and \ref{eqn:g2_gauss}, for the first and second orders, respectively (red dashed line).}
    \label{fig:sim_comp_1}
\end{figure}

\begin{figure}[h]
    \centering
    \includegraphics[width=1\textwidth]{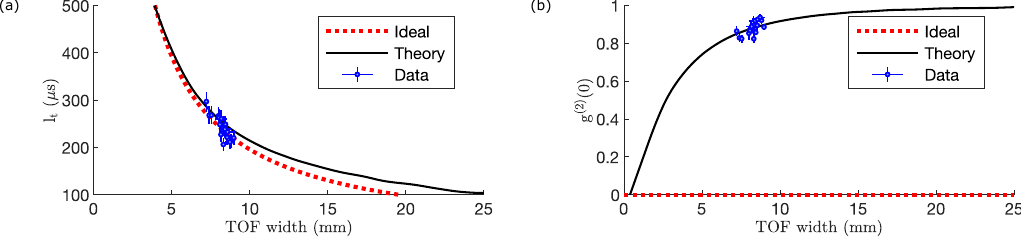}
    \caption{The experimentally measured temporal correlation length $l_t$ (a) and correlation amplitude $g^{(2)}(0)$ (b) for various time-of-flight widths of the \HeThreeStar{} cloud compared to the theoretically ideal value (red dashed line) and the results of our Monte Carlo simulation (black solid line).}
    \label{fig:sim_comp_2}
\end{figure}

While the results from Sec.~\ref{sec:wick} and \ref{sec:exact} indicate the expected correlation length and amplitude at a given order for a perfect detection system, experimental measurements will typically yield reduced amplitudes, due to a variety of effects. To understand the most relevant effects, we will briefly describe how the $n$th order correlation function is computed for this particular work. We first compute the spatial (both in $x$ and $y$ axes) and temporal differences between all pairs of atom detections of a given atomic species. For the $n$th order correlation function, we consider all $n$-tuples of particles, for example, pairs in second order and triplets in third order, whose spatial difference between all pairs of particles in the tuple are less than a given threshold $\Delta x$ and $\Delta y$ for the $x$ and $y$ axes respectively. This is practically equivalent to integrating the full correlation function over these dimensions. Hence, we expect improved correlation amplitudes for decreasing $\Delta x$ and $\Delta y$ at the cost of signal-to-noise performance. We then take all the temporal difference $\tau_1,\ldots,\tau_{n-1}$, repeating for all counts in a given experimental realisation. These differences can then be histogrammed to obtain the unnormalised $n$-th order correlation function along the time dimension. The unormalised distribution is averaged over many experimental realisations to improve signal-to-noise. To normalise the distribution, we repeat the above procedure, however this time for counts recorded in different experimental realisations, which cannot possibly interfere and hence should contain no underlying correlation allowing us to determine the bulk distribution. Note that detection efficiency does not affect the normalised correlation function, as it proportionally affects the numerator and the denominator by the same amount. From this, we see that the strongest deviation from the ideal case comes from the effective integration over the $x$ and $y$ axes due to the finite bin size. There is also a slight integration over the time axis due to again finite bin size. This effect compounds at higher orders, as for each new particle added we integrate over another set of axes. Less noticeable, but still significant effects are the resolution of the detector and the dark, or background, count rate of the detector. 
\\ \\
To account for these effects, we can perform a Monte Carlo simulation of our full measurement process, as described above, assuming the underlying true distribution follows the relations in Sec.~\ref{sec:wick} and \ref{sec:exact} and with the various imperfections included. As our detection parameters, such as resolution and dark count rate, are fixed, and the ideal correlation amplitude is always zero for any order, the only relevant variable in our simulation is the correlation length. The correlation length of the trapped gas, as derived in Sec.~\ref{sec:exact}, is given by $l_i = \frac{\hbar}{2s_i} \gamma(\xi)$. As mentioned above, in practice, we experimentally measure the correlations in a ballistically expanding cloud. Hence, we rescale the momentum space correlation length to give the experimentally measured TOF correlation lengths,
\begin{table}[t]
    \centering
    \begin{tabular}{lcc}
    \toprule
    \toprule
        Parameter & Value & Reference \\
        \hline
        detector resolution ($t,\, x,\, y$) & $\Big($$3$~$\mu$s, $120$~$\mu$m, $120$~$\mu$m$\Big)$ & \cite{s1,hbt2}  \\
        Flight time ($t_{TOF}$) & 0.417~s & \\
        Trapping frequency $(\omega_{x},\omega_{y},\omega_{z})$ \, \, & $2\pi \Big(58(3),694(1),701(2)\Big)$ Hz & \cite{Thomas2023, Henson2022} \\
        Dark count rate & $0.046$~mm$^{-2}$s$^{-1}$ & \\
        Bin size $(\Delta t, \, \Delta x, \, \Delta y)$ &
         $\Big($$100$~$\mu$s, $130 \, \mu$m, $420 \, \mu$m$\Big)$ & \\
        \bottomrule
    \end{tabular}
    \caption{Detection and analysis parameters used in Monte Carlo simulation. Analysis parameters are the same as those used to analyze the experimental data.}
    \label{tab:monte_carlo}
\end{table}

\begin{align}
    l^{TOF}_i=\frac{\hbar t_{TOF} \gamma(\xi)}{2m s_i},  \label{eqn:corr_length_sp}
\end{align}
where $t_{TOF}$ is the flight time, and now $i\in (t,x,y)$. For the $t$-axis, as we measure arrival time rather than a location, the TOF scaling is slightly different. If the expansion of the cloud is negligible compared to the fall distance, we can consider a correlation time of
\begin{align}
    l_t&=\frac{l^{TOF}_z}{g t_{TOF}} = \frac{\hbar \gamma(\xi)}{2m s_z g},  \label{eqn:corr_length_t}
\end{align}
where $g$ is the acceleration due to gravity. 
\\ \\
We can now test the validity of our model by varying the TOF correlation length and comparing the results to the measured amplitude and correlation length. In practice, we vary the correlation length by changing the temperature of the trapped cloud. However, as it is difficult to ascertain an accurate measure of temperature from the TOF profile of the \HeThreeStar{} cloud alone, we instead use the (isometric) time-of-flight width of the cloud, given by $\sigma_{TOF}=\frac{t_{TOF}}{\gamma(\xi)} \sqrt{\frac{k_B T}{m}}$, as a proxy for temperature. From Eqn.~\ref{eqn:corr_length_sp}, we can see that $l^{TOF}_i=\frac{\hbar  \omega_i}{2m \sigma_{TOF}} t_{TOF}^2$, and thus, as $\omega_i$ is fixed $\sigma_{TOF}$ alone is a sufficient initial condition for our Monte Carlo simulation and can be directly compared to experiment.
\\ \\
The relevant experimental parameters used in our simulation are listed in Tab.~\ref{tab:monte_carlo}. The resultant $g^{(1)}$ and $g^{(2)}$ distributions for correlation lengths of $(l_t, \, l_x,\, l_y)$=$(244.6\,\mu\text{s},\, 83.0 \,\mu\text{m},\, 996.6 \,\mu\text{m})$ ($\sigma_{TOF}=8$~mm) with and without the experimental imperfections are shown in Fig.~\ref{fig:sim_comp_1}. The predicted values for the experimentally measured amplitude and correlation length for various cloud sizes are shown in Fig.~\ref{fig:sim_comp_2}, showing good agreement between theory and experiment and demonstrating that it is crucial to include these realistic imperfections in order to obtain accurate predictions.

\end{document}